\documentclass[aps,prl,twocolumn,showpacs,preprintnumbers,amsmath,amssymb,superscriptaddress]{revtex4-1}

 \usepackage{graphicx}

\begin{document}

\title{Amplifying single impurities immersed in a gas of ultracold atoms}
\author{B. Olmos}
\affiliation{Midlands Ultracold Atom Research Centre (MUARC). School of Physics and Astronomy, University of Nottingham, Nottingham, NG7 2RD, United Kingdom}
\author{W. Li}
\affiliation{Midlands Ultracold Atom Research Centre (MUARC). School of Physics and Astronomy, University of Nottingham, Nottingham, NG7 2RD, United Kingdom}
\author{S. Hofferberth}
\affiliation{5. Physikalisches Institut, Universit\"at Stuttgart, Pfaffenwaldring 57, 70569 Stuttgart, Germany}
\author{I. Lesanovsky}
\affiliation{Midlands Ultracold Atom Research Centre (MUARC). School of Physics and Astronomy, University of Nottingham, Nottingham, NG7 2RD, United Kingdom}

\begin{abstract}
We present a method for amplifying a single or scattered impurities immersed in a background gas of ultracold atoms so that they can be optically imaged and spatially resolved. Our approach relies on a Raman transfer between two stable atomic hyperfine states that is conditioned on the presence of an impurity atom. The amplification is based on the strong interaction among atoms excited to Rydberg states. We perform a detailed analytical study of the performance of the proposed scheme with particular emphasis on the influence of inevitable many-body effects.
\end{abstract}

\pacs{}

\maketitle
\begin{figure}
\centering
\includegraphics[width=\columnwidth]{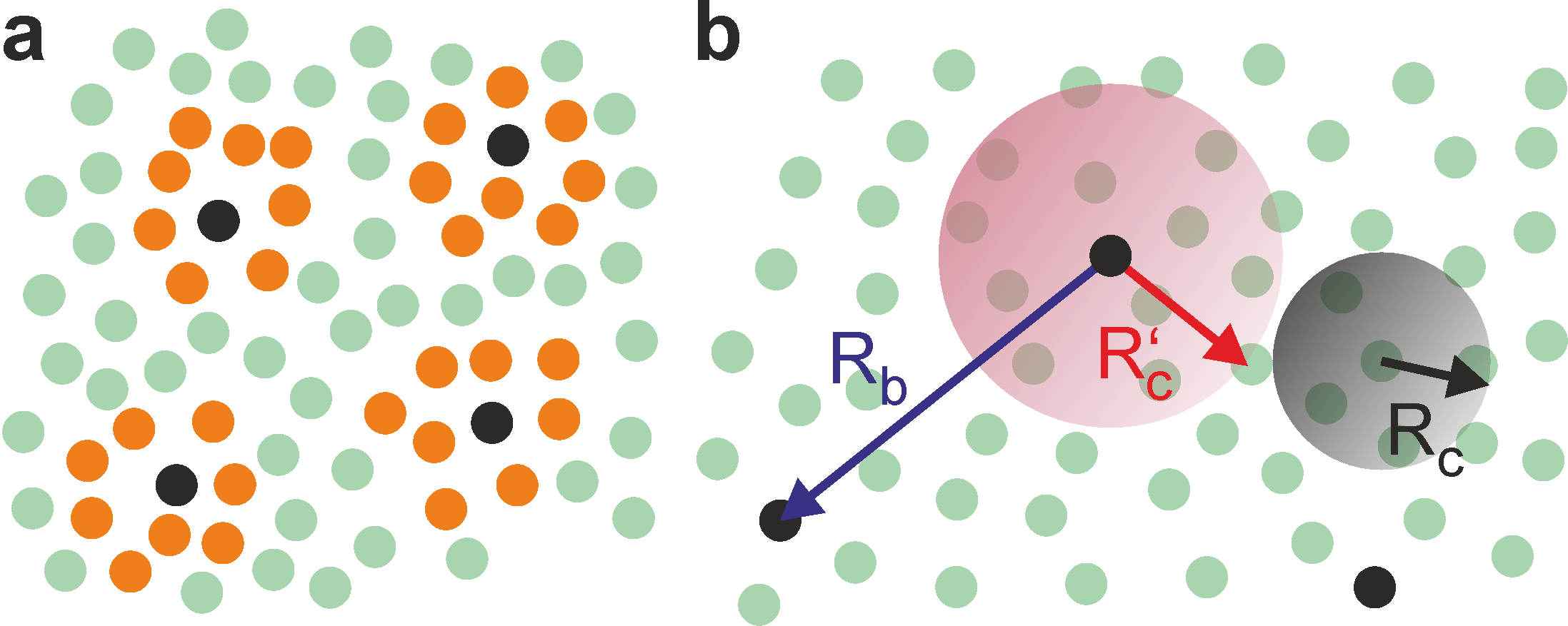}
\caption{\textbf{a:} After the amplification step the atoms of the background gas have changed their internal state [from $\left|A\right>$ (green) to $\left|B\right>$ (orange)] in the vicinity of an impurity (black). State selective absorption imaging of the background gas then permits the spatial discrimination of the impurity atoms. \textbf{b:} Length scales. The mean distance $R_\mathrm{b}$ between the impurity atoms is much larger than the mean interparticle distance in the uniform background gas. Atoms within a sphere with radius $R^\prime_\mathrm{c}$ change their internal state. Background atoms located within $R_\mathrm{c}$ are interacting (see text for explanation).}\label{fig:gas}
\end{figure}

Impurities inside interacting quantum fluids provide a variety of interesting physical effects in fields such as high-T$_\mathrm{c}$ superconductivity or plasma physics. Recent experimental advances in creating and controlling impurities in ultracold quantum gases have demonstrated that these systems are ideally suited for studying impurity phenomena such as transport in strongly correlated 1D-Bose gases \cite{Palzer09}, polaron physics in highly imbalanced Fermi-gas mixtures \cite{Sommer11}, or Rydberg-excitation saturation in Bose-Einstein condensates \cite{Heidemann08}. Since detection of single (neutral) impurities in a quantum gas poses severe experimental challenges, these experiments all rely on measuring either ensemble averages or effects of the impurities on the bulk host medium. Here, we develop a novel single shot detection scheme for single impurity atoms, based on strong amplification of the detectable signal by converting the impurity into an internal state change of a large number of nearby background atoms as shown in Fig. \ref{fig:gas}a. Our scheme is applicable to any form of neutral impurity in an ultracold gas and can be implemented with standard absorption-imaging techniques available in virtually every existing experiment, e.g. one possible application is the visualization of Rydberg crystals in a Bose gas \cite{Pohl10}.

In the experimental situation we have in mind the background gas is initially polarized with all atoms being in a stable hyperfine state $\left|A\right>$. The average distance $R_\mathrm{b}$ between the impurities, which can be atoms of the same species in other internal states or a different atomic species altogether, is much larger than the average distance between background atoms. As illustrated in Fig. \ref{fig:gas}b, our amplification scheme performs a conversion of background atoms from $\left|A\right>$ to a second hyperfine state $\left|B\right>$ \emph{only within a sphere} of radius $R^\prime_\mathrm{c}$ centered at the impurity. These atoms can be detected via state selective absorption imaging in a single shot and provided that $R_\mathrm{b}> 2R^\prime_\mathrm{c}$ a discrimination of the positions of the impurities can be achieved. We show that our scheme can be directly implemented in current experiments with ultracold Rubidium gases yielding amplification factors of $\sim 10-100$ atoms per impurity.

\begin{figure}
\centering
\includegraphics[width=\columnwidth]{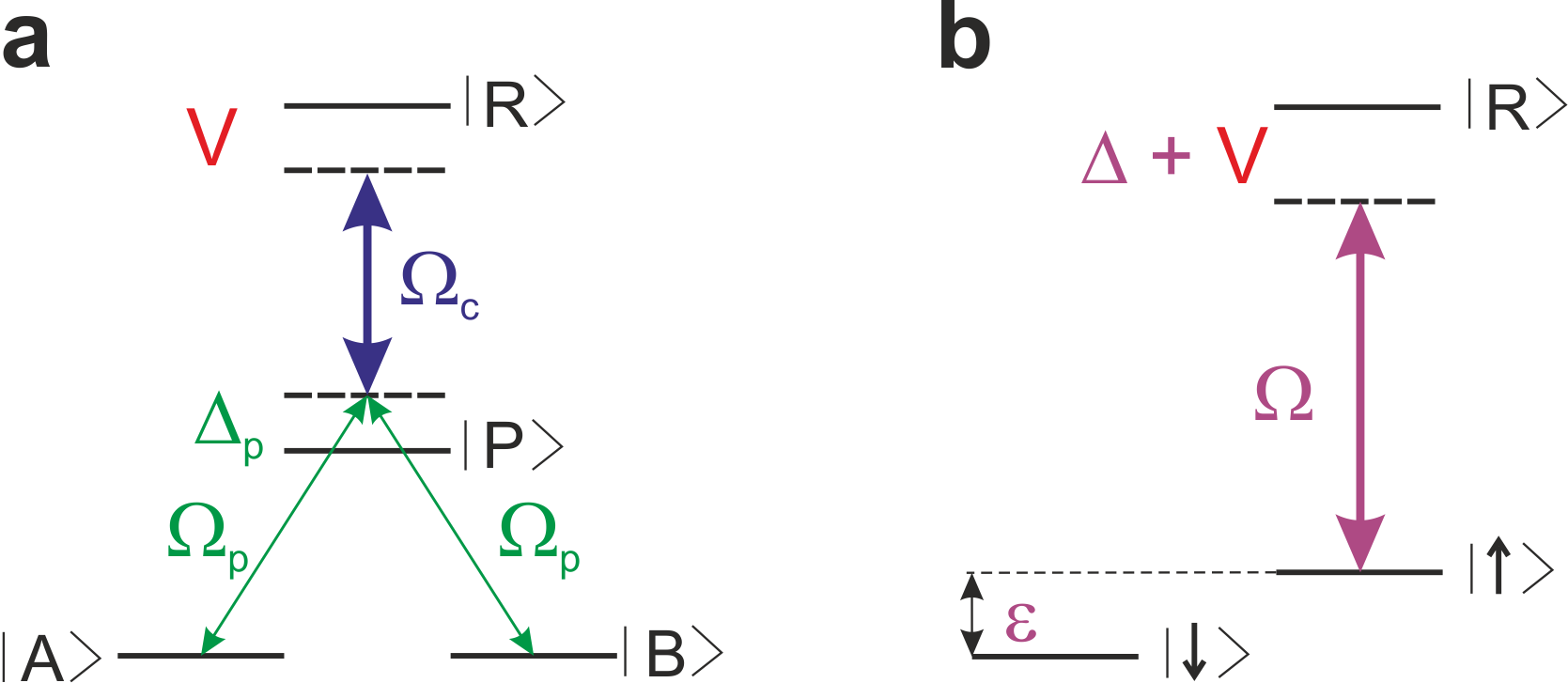}
\caption{Internal level scheme of the atoms of the background gas. \textbf{a}: The two hyperfine ground states $\left|A\right>$ and $\left|B\right>$ are coupled off-resonantly to the intermediate $\left|P\right>$ state by means of a laser (detuning $\Delta_\mathrm{p}$, Rabi frequency $\Omega_\mathrm{p}$). A strong laser with Rabi frequency $\Omega_\mathrm{c}$ couples $\left|P\right>$ to a Rydberg state $\left|R\right>$. The interaction between the impurity and the background atom $V$ gives rise to an effective detuning of the Rydberg level. \textbf{b}: The scheme can be mapped into a two-level system formed by $\left|\uparrow\right>=(1/\sqrt{2})\left[\left|A\right>+\left|B\right>\right]$ (with energy $\epsilon$) and the Rydberg level, coupled with an effective Rabi frequency $\Omega$ and detuning $\Delta+V$ and an uncoupled level $\left|\downarrow\right>=(1/\sqrt{2})\left[\left|A\right>-\left|B\right>\right]$.}
\label{system}
\end{figure}

We start from a setting where all impurity atoms have been electronically excited to a Rydberg $n^\prime S$-state. Such situation can be easily achieved by a resonant $\pi$-pulse from the ground state. As an example, one can consider the case of strongly interacting Rydberg atoms in a cloud of ultracold Rubidium atoms, where the typical distance between impurities is given by the so-called Rydberg blockade radius \cite{Lukin01,Saffman10-2} which can be on the order of ten micrometers \cite{Tong04,Singer04,Pillet06,Urban09,Gaetan09}. To effectuate the conditional transfer between the hyperfine states of each background atom we use the atomic levels depicted in Fig. \ref{system}a. This scheme has been thoroughly discussed in Ref. \cite{Mueller09} in the context of a mesoscopic quantum gate. The level $\left|P\right>$ is a metastable low-lying $P$-state over which the hyperfine states of interest ($\left|A\right>$ and $\left|B\right>$) are off-resonantly coupled (detuning $\Delta_\mathrm{p}$, Rabi frequency $\Omega_\mathrm{p}$). The state $\left|P\right>$ is furthermore coupled to a Rydberg $n S$ state $\left|R\right>$ with a laser of Rabi frequency $\Omega_\mathrm{c}$ whose frequency is chosen such that the states $\left|A\right>$ and $\left|R\right>$ are in two-photon resonance. The interaction between the Rydberg state and the impurity, denoted by $V$, acts as an effective detuning for the $\left|R\right>$ state. Thus the Hamiltonian of a single background atom is given by $H_\mathrm{scheme}=\Omega_\mathrm{p}\left[\left|A\right>\!\left<P\right|+\left|B\right>\!\left<P\right|+\mathrm{h.c.}\right]+
\Omega_\mathrm{c}\left[\left|R\right>\!\left<P\right|+\mathrm{h.c.}\right]-\Delta_\mathrm{p}\left|P\right>\!\left<P\right|+ V\left|R\right>\!\left<R\right|$, with $\hbar=1$.
Upon adiabatic elimination of the far-detuned $P$-state ($|\Delta_\mathrm{p}|\gg \Omega_\mathrm{p},\Omega_\mathrm{c}$) and the introduction of the superposition states $\left|\uparrow\right>=(1/\sqrt{2})\left[\left|A\right>+\left|B\right>\right]$ and $\left|\downarrow\right>=(1/\sqrt{2})\left[\left|A\right>-\left|B\right>\right]$ the system reduces to a two-level system plus an uncoupled third level (see Fig.\ref{system}b). The corresponding Hamiltonian reads
\begin{equation*}
H_\mathrm{ad}=\Omega\left[\left|\uparrow\right>\!\left<R\right|+\mathrm{h.c.}\right]+\left(\Delta+V\right)\left|R\right>\!\left<R\right|-
\epsilon\left|\downarrow\right>\!\left<\downarrow\right|
\end{equation*}
with the (effective) Rabi frequency and detuning $\Omega=\sqrt{2}\Omega_\mathrm{p}\Omega_\mathrm{c}/\Delta_\mathrm{p}$ and
$\Delta=(\Omega^2_\mathrm{c}-2\Omega^2_\mathrm{p})/\Delta_\mathrm{p}$, respectively and where $\epsilon=2\Omega^2_\mathrm{p}/\Delta_\mathrm{p}$ is the energy of the $\left|\uparrow\right>$-state. This scheme formally corresponds to an off-resonant Raman coupling between $\left|A\right>$ and $\left|B\right>$ via the Rydberg state. Note however, that here $\Omega$, $\Delta$ and $\epsilon$ are not independent from one another.

To illustrate the mechanism behind the amplifier, we consider a single impurity atom localized at the origin of the coordinate system. The interaction energy between this impurity (excited to the Rydberg state $n^\prime S$) and the $k$-th atom of the background gas located at a distance $|\mathbf{R}_k|=R_k$ is given by $V_k=C^\prime_6/R^6_k$. Here $C^\prime_6$ is the van-der-Waals coefficient for the $n^\prime S - nS$ interaction. This interaction energy together with the detuning $\Delta$ defines a spatially varying effective detuning of the Rydberg state of the $k$-th background atom: $\Delta_k=\Delta+V_k=\Delta(1+{R^\prime_\mathrm{c}}^6/R^6_k)$ where we have defined the radius $R^\prime_\mathrm{c}=(C^\prime_6/\Delta)^{1/6}$. As we will show, this spatially dependent energy shift is the origin of the conditional transfer between the hyperfine states of the background atoms.
For $\Omega_\mathrm{c}\gg\Omega_\mathrm{p}$, which we assume throughout, the probability to excite a Rydberg state is $\sim(\Omega/\Delta)^2\ll 1$. We can thus adiabatically eliminate the state $\left|R\right>$ and find that each background atom evolves effectively under the Hamiltonian $H_\mathrm{\uparrow\downarrow}=-\epsilon\left[\frac{R_k^6}{{R^\prime_\mathrm{c}}^6+R_k^6} \left|\uparrow\right>\!\left<\uparrow\right|+\left|\downarrow\right>\!\left<\downarrow\right|\right]$. We switch on the effective laser coupling (Rabi frequency $\Omega$) for a time $\tau_\pi=\pi\Delta_\mathrm{p}/2\Omega_\mathrm{p}^2=\pi/\epsilon$ ($\pi$-pulse). This results in a conditional transfer between the hyperfine states as a consequence of the spatially dependent phase difference that is accumulated between the states $\left|\uparrow\right>$ and $\left|\downarrow\right>$ under the action of $H_\mathrm{\uparrow\downarrow}$: If the background atom is close to the impurity ($R_k\rightarrow 0$), the coupling realizes the transition $\left|A\right>\rightarrow \left|B\right>$, i.e. the atoms of the background gas change their state since $\left|\downarrow\right>$ acquires a phase of $\pi$ relative to $\left|\uparrow\right>$. Far from the impurity ($R_k\gg R^\prime_\mathrm{c}$) the acquired phase difference is zero and hence all atoms remain in the state $\left|A\right>$.

While the mechanism is intuitive for a single background atom, its generalization to the many-body case is not straightforward. This is due to the fact that the atoms forming the background gas are coupled to Rydberg states during the amplification procedure which then leads to an interaction among themselves. The importance of this interaction has been impressively demonstrated in recent experiments \cite{Pritchard10} and theoretical work \cite{Ates11,Petrosyan11,Lukin11,Sevincli11} that investigate electromagnetic transparency in Rydberg gases. In the following we give a detailed account of these many-body effects. The interaction energy between two background atoms in the Rydberg state is given by $V_{km}=C_6/|\mathbf{R}_k-\mathbf{R}_m|^6$, where $\mathbf{R}_k$ and $\mathbf{R}_m$ are their respective spatial positions and $C_6$ is the van-der-Waals coefficient for the $nS-nS$ interaction. We proceed by adiabatically eliminating all many-body states that contain one or more Rydberg excitations \cite{Pupillo10,Henkel10}. After this elimination, we can rewrite the Hamiltonian in terms of the spin operator $S_z^{(k)}$ that acts on the background atom located at $\mathbf{R}_k$ and we are left with the effective Hamiltonian
\begin{align}
  H=\sum_k \epsilon_k S^{(k)}_z+\sum_{k,m}M_{km}S^{(k)}_z + \sum_{k,m}M_{km}S^{(k)}_zS^{(m)}_z\label{eq:ad_elim_Hamiltonian}
\end{align}
with $\epsilon_k=\epsilon-\frac{\Omega^2}{\Delta_k}$ and
\begin{eqnarray*}
  M_{km}=\Omega^4\frac{V_{km}\left(\Delta_k+\Delta_m\right)}{2\Delta_k^2\Delta_m^2\left(\Delta_k+\Delta_m+V_{km}\right)}.
\end{eqnarray*}
This Hamiltonian is valid for atomic densities $\rho$ that obey $\rho\, (C_6/\Delta)^{D/6} (\Omega/\Delta)^2<1$, where the two-body potential dominates the many-body interactions \cite{Honer10}, with $D$ being the dimension of the system. The Hamiltonian (\ref{eq:ad_elim_Hamiltonian}) is interpreted as follows: The first term is the sum of independent atoms that undergo a rotation about the spin $z$-axis depending on the distance to the impurity (similar to the Hamiltonian $H_\mathrm{\uparrow\downarrow}$). The second one effectuates an extra rotation about the $z$-axis due to the interaction between the Rydberg states of the background atoms. The third term is a spin-spin interaction. We first focus our discussion on the first two terms, and show later that the third term can be understood as a squeezing interaction.

We employ a continuum representation for the Hamiltonian (\ref{eq:ad_elim_Hamiltonian}) such that $\sum_k\rightarrow \rho \int d^D\mathbf{R}$. The double sum in the second term then turns into a double integral over the spatial coordinates $\mathbf{R}_1$ and $\mathbf{R}_2$. To simplify the Hamiltonian we perform a coarse graining procedure in which we describe all atoms located at distance $R$ from the impurity in terms of a spin density $S_z(R)$, i.e. $\sum_{R_k=R}S^k_z= dR R^{D-1}S_z(R)\int\!\!d\mathbf{\Omega}_{D-1}$ \footnote{$\int\!\!d\mathbf{\Omega}_{D-1}$ represents the integral over the solid angle in $D$ dimensions. Note that $\int\!\!d\mathbf{\Omega}_{0}=2$.}. This is motivated by the fact that in a background gas with uniform density the interaction energy of each atom of the gas can only depend on its distance $R$ to the impurity. We introduce the relative and center of mass coordinates $\mathbf{r}=\mathbf{R}_1-\mathbf{R}_2$ and $\mathbf{R}=(\mathbf{R}_1+\mathbf{R}_2)/2$, respectively, integrate over the relative coordinates and average over the solid angle of the center of mass. This leads us to
\begin{eqnarray*}
  H=\int \!\!dR R^{D-1}S_z(R)\int d\mathbf{\Omega}_{D-1} \left[f_1(D,R)+f_2(D,R)\right]
\end{eqnarray*}
with
\begin{equation*}
  f_1(D,R)=\epsilon-\frac{\Omega^2}{\Delta}\frac{R^6}{R^6+{R^\prime_\mathrm{c}}^6}
\end{equation*}
and
\begin{eqnarray*}
  f_2(D,R)&=&\frac{\Omega^4}{\Delta^3}\frac{\rho}{2\int d\mathbf{\Omega}_{D-1}} \int d^{D}\mathbf{r}\!\!\int d\mathbf{\Omega}_{D-1}\\
  &&\times\frac{\alpha x(2+x_1+x_2)}{(1+x_1)^2(1+x_2)^2(2+x_1+x_2+\alpha x)},
\end{eqnarray*}
where $x_k={R_\mathrm{c}^\prime}^6/R_k^6$, $x={R_\mathrm{c}^\prime}^6/|\mathbf{r}|^6$ and $\alpha=C_6/C^\prime_6$. One can rewrite the previous expression as
\begin{eqnarray}\label{eq:interaction}
  f_2(D,R)=\frac{\Omega^4}{\Delta^3}\xi_D\rho R_\mathrm{c}^D g_D(R,\alpha),
\end{eqnarray}
where we have defined the radius $R_\mathrm{c}=(C_6/\Delta)^{1/6}$ and where $\xi_D$ is a numerical factor that depends on the dimension $D$ of the system: $\xi_1=\frac{2\pi}{3\,2^{1/6}}$, $\xi_2=\frac{2\pi^2}{3\sqrt{3}\,2^{1/3}}$ and $\xi_3=\frac{2\pi^2}{3\,2^{1/2}}$. The function $g_D(R,\alpha)$ is $0$ for $R\rightarrow 0$ and $1$ for $R\gg R^\prime_\mathrm{c}$. It only weakly depends on the dimension and $\alpha$, and it is well approximated by $g_D(\alpha,R)\approx R^7/(R^7+10 {R^\prime_\mathrm{c}}^7)$, provided that $0\leq \alpha\leq 10$.

Next we discuss the actual dynamics of the amplification procedure in the many-body picture. Initially the background gas is prepared in the state $\left|A\right>^N$, where $N=\int d\mathbf{\Omega}_{D-1} \int dR R^{D-1}\rho$. The time-evolution of this state can be studied for each shell of radius $R$ independently. The mean density of $\left|B\right>$-atoms in such a shell is given by
\begin{equation}
  \rho_\mathrm{B}(R,t)=\frac{\rho}{2}\left\{1-\cos\left[t \left(f_1(D,R)+f_2(D,R)\right)\right]\right\}\label{eq:density}.
\end{equation}
We are interested in the state of the background atoms after a $\pi$-pulse of duration $\tau_\pi=\pi/\epsilon$. Close to the impurity ($R\rightarrow 0$) we find $\rho_\mathrm{B}(0,\tau_\pi)=\rho$ while for $R\gg R^\prime_\mathrm{c}$ the density is approximately
\begin{eqnarray}
  \rho_\mathrm{B}(\infty,\tau_\pi)\approx\rho\pi^2\frac{\Omega_\mathrm{p}^4}{\Omega_\mathrm{c}^4} \left(\xi_D\rho R_\mathrm{c}^D\right)^2,\label{eq:density_infinity}
\end{eqnarray}
where we have expanded in $f_2(D,R)\tau_\pi\ll1$. This background of $\left|B\right>$-atoms is caused by a mean field energy shift due to virtually excited background atoms that introduces a small phase difference between the states $\left|\uparrow\right>$ and $\left|\downarrow\right>$. \emph{This clearly illustrates the importance of the interaction between the Rydberg states of the background atoms.} This effect can be minimized by choosing either Rydberg states in which the background atoms interact weakly (small $R_\mathrm{c}$), by decreasing the ratio $\Omega_\mathrm{p}/\Omega_\mathrm{c}$ or by reducing the density $\rho$. The latter is not desirable as this reduces the amplification factor, i.e. the total number of $\left|B\right>$-atoms produced by an impurity which is given by
\begin{eqnarray}\nonumber
  N_\mathrm{B}&=&\int d\mathbf{\Omega}_{D-1}\int dR R^{D-1}\left[\rho_\mathrm{B}(R,\tau_\pi)-\rho_\mathrm{B}(\infty,\tau_\pi)\right]\\
  &\approx &\rho\,{R^\prime_\mathrm{c}}^D\!\!\int\!\! d\mathbf{\Omega}_{D-1}\left[\gamma_D + \lambda_D\pi\frac{\Omega_p^2}{\Omega_c^2}\xi_D \rho\,R_\mathrm{c}^D\right]\label{eq:NB}
\end{eqnarray}
where the second line is again the result of an expansion in $f_2(D,R)\tau_\pi\ll1$. The parameters $\gamma_D=\{1.02, 0.53, 0.39\}$ and $\lambda_D=\{0.16, 0.24, 0.41\}$ have been calculated with the analytic approximation to $g_D(\alpha,R)$.
\begin{figure}
\centering
\includegraphics[width=0.8\columnwidth]{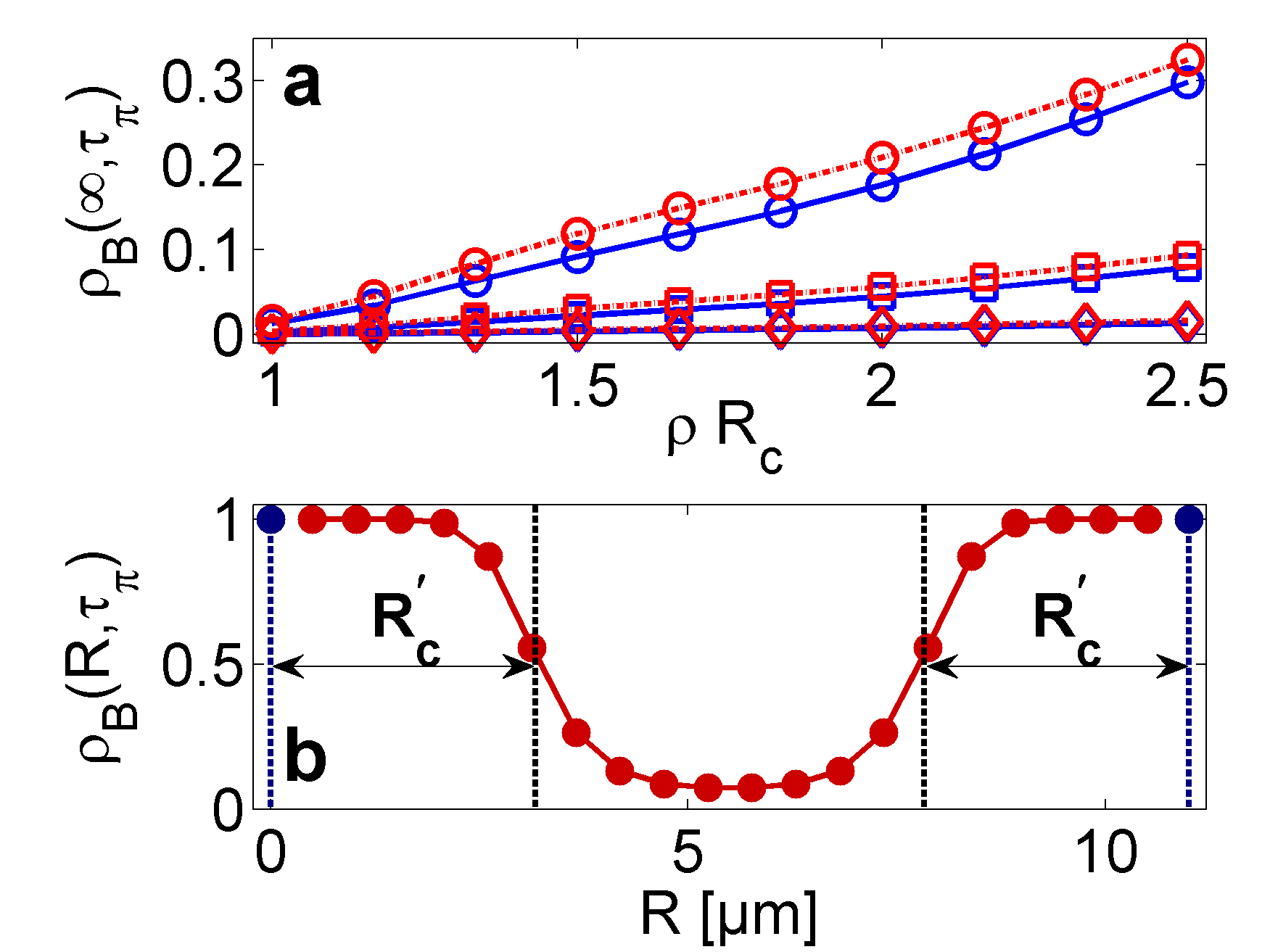}
\caption{\textbf{a:} Background density of $\left|B\right>$ atoms after the Raman transfer versus the parameter $\rho R_\mathrm{c}$. The data shown correspond to numerical simulations of a one-dimensional lattice gas of $N=20$ atoms. For each value of $\Omega_\mathrm{p}/\Omega_\mathrm{c}=0.1$ (diamonds) $0.15$ (squares) and $0.2$ (circles), two cases are depicted: the data using the complete Hamiltonian (\ref{eq:ad_elim_Hamiltonian}) (red dotted lines) and without taking into account the third term proportional to $S_z^{(k)}S_z^{(m)}$ (blue solid lines). \textbf{b:} Density of atoms in the $\left|B\right>$ state for the same system, with $\Omega_\mathrm{p}/\Omega_\mathrm{c}=0.1$ and one impurity atom at each end of the lattice (blue dots). The values of the parameters considered are: principal quantum numbers of the background and impurity atoms $n=39$ and $n^\prime=62$, respectively; $\Omega_\mathrm{p}/(2\pi)=10\,$MHz; $\Omega_\mathrm{c}/(2\pi)=100\,$MHz; $\Delta_\mathrm{p}/(2\pi)=1\,$GHz; $\rho=2\,\mu \mathrm{m}^{-1}$. The impurity is amplified by a factor $\sim 10$.}
\label{fig:contrast}
\end{figure}

Let us discuss the role of the last term (spin-spin interaction) of the Hamiltonian (\ref{eq:ad_elim_Hamiltonian}) which we have omitted so far. Its effect can be estimated as follows: Each atom of the background gas is interacting with other background atoms in the volume $\sim \xi_DR_\mathrm{c}^Dg_D(R,\alpha)$. Within this volume (see Fig. \ref{fig:gas}b) only one of the background atoms can be (virtually) excited. We consider now a picture in which the space is divided into independent regions of the latter volume. This allows us to rewrite the last term of Hamiltonian (\ref{eq:ad_elim_Hamiltonian}) as a spin squeezing interaction \cite{Bouchoule02} that in general affects the transfer of background atoms from $\left|A\right>$ to $\left|B\right>$. This effect can be approximately accounted for by multiplying the factor $\cos^{\xi_D\rho R_\mathrm{c}^Dg_D(R,\alpha)-1}\left(\frac{\Omega^4}{\Delta^3}t\right)$ in front of the $\cos$-term in eq. (\ref{eq:density}) (see Ref. \cite{Kitigawa93}). Due to the functional form of $g_D(R,\alpha)$, this factor is close to one near the impurity and atoms are transferred between the hyperfine states as anticipated. The effect for $R\gg R^\prime_\mathrm{c}$  is demonstrated in Fig. \ref{fig:contrast}a which shows results obtained from numerically solving Hamiltonian (\ref{eq:ad_elim_Hamiltonian}) on a 1D background lattice gas of $20$ Rubidium atoms. As predicted the background density of $\left|B\right>$-atoms $\rho_\mathrm{B}(\infty,\tau_\pi)$ grows with increasing $\rho R_\mathrm{c}$. The effect of the spin-spin interaction is marginal. Fig. \ref{fig:contrast}b shows that two impurities at a distance $R_\mathrm{b}\sim 10\,\mu$m can be resolved by optically imaging atoms in the state $\left|B\right>$. The agreement with the analytical prediction is excellent.

\begin{figure}
\centering
\includegraphics[width=0.8\columnwidth]{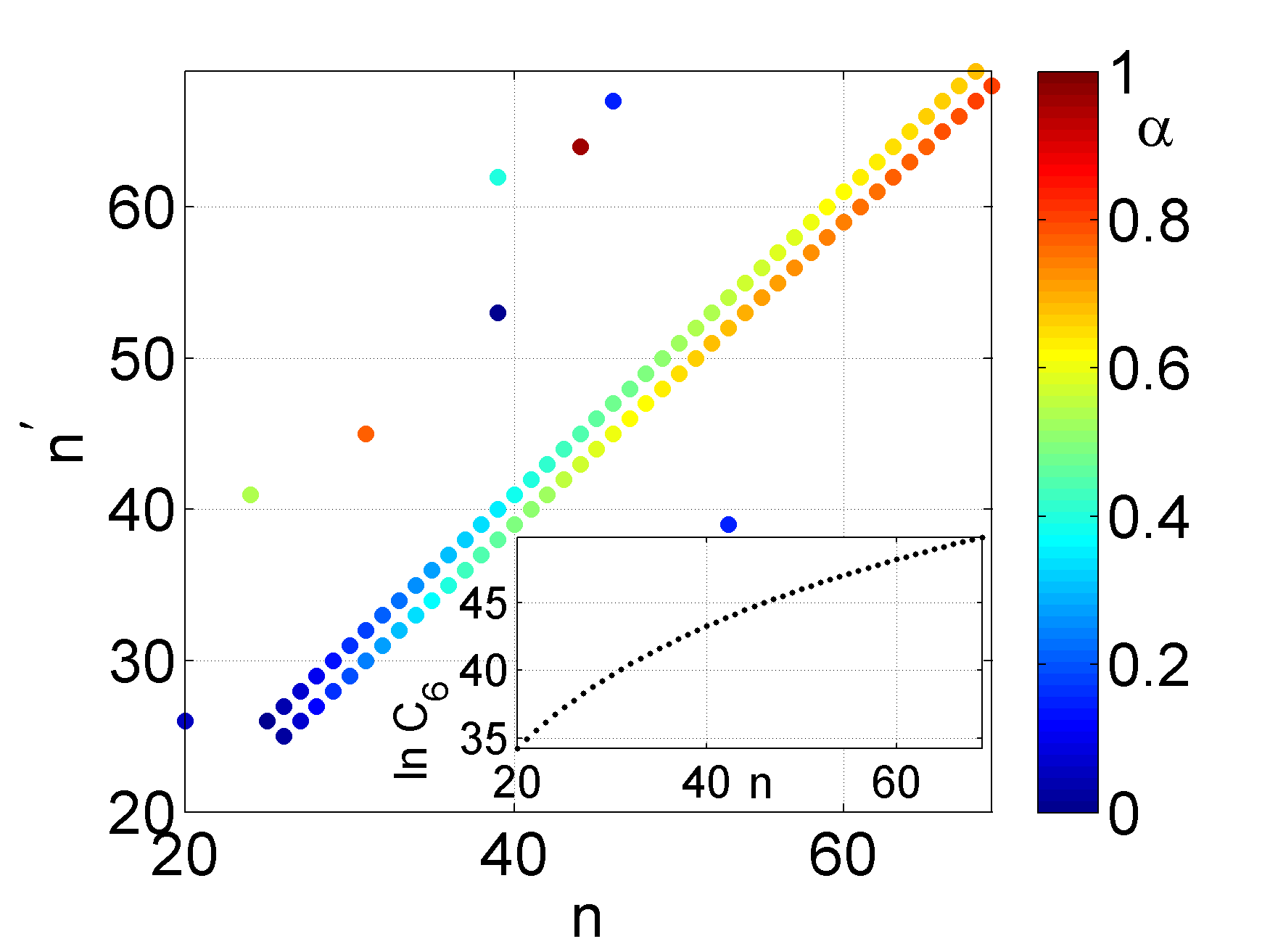}
\caption{Ratio $\alpha=C_6/C_6^\prime$ for pairs of Rydberg S-states of Rubidium with principal quantum numbers $(n,n^\prime)$ that accomplish $0<\alpha<1$. In the inset, the value of the $C_6$ in atomic units calculated by using second order perturbation theory as in Ref. \cite{Singer05}.}
\label{fig:C6}
\end{figure}

We now present experimental parameters for the implementation of our scheme in current experiments in ultracold gases of Rubidium. In order to maximize the contrast, we need the background atoms to interact strongly with the impurity atoms and weakly among themselves, i.e. $\alpha=C_6/C_6^\prime<1$. Moreover, the sign of the two interactions must be positive in order to avoid unwanted resonances. This is accomplished by only a few combinations of principal quantum numbers $(n,n^\prime)$ as shown in Fig. \ref{fig:C6}.
To avoid loss of atoms from the intermediate $\left|P\right>$ state, the duration $\tau_\pi$ of the Raman transfer must be shorter than the lifetime of the state ($\tau_p\approx 27\,\mathrm{ns}$ for Rubidium $5P$ state) divided by the probability of exciting the state (given by $\Omega_\mathrm{p}^2/\Delta_\mathrm{p}^2$). This yields a lower limit for the detuning $\Delta_\mathrm{p}/(2\pi)\gg 50\,$MHz. Moreover, $\tau_\pi$ has to be shorter than the lifetime of the Rydberg state of the impurity, typically on the order $T_\mathrm{R}\approx 100\,\mu$s. This sets a minimum value for $\Omega_\mathrm{p}/(2\pi)\gg \sqrt{\pi(\Delta_\mathrm{p}/(2\pi))/(2T_\mathrm{R})}$, that will depend on the particular choice of the impurity state. Finally, $\Omega_\mathrm{p}/\Omega_\mathrm{c}\ll1$ for the Hamiltonian (\ref{eq:ad_elim_Hamiltonian}) to be valid. Taking all this into account the choice of parameters $\Delta_\mathrm{p}/(2\pi)=1\,$GHz, $\Omega_\mathrm{p}/(2\pi)\approx 10\,$MHz and $\Omega_\mathrm{c}/(2\pi)\approx 100\,$MHz fixes the time of the pulse to $\tau_\pi\approx 15\,\mu$s, much shorter than the lifetime of any Rydberg $n^\prime$-S state with $n^\prime>40$. This time is also much shorter than the typical timescale of the atomic motion (milliseconds), so that we can consider the gas to be \emph{frozen} \cite{Mourachko98,Anderson98} for the duration of the transfer. The radius $R^\prime_\mathrm{c}$ of the sphere inside which atoms are transferred to the imaging state can be tuned depending on the mean impurity distance $R_\mathrm{b}$ and available imaging resolution. This can be done by selecting an appropriate pair of Rydberg states, to which the impurities and background atoms are excited. As an example, in typical experiments studying Rydberg-blockade effects, $R_\mathrm{b}$ is on the order of $10\,\mu$m \cite{Pohl10}, while imaging resolutions are $\le 3\,\mu$m. For $R^\prime_\mathrm{c} = 3\,\mu$m and atomic densities $\rho\sim 10^{11}\,\mathrm{cm}^{-3}$, the number of atoms in $\left|B\right>$ around an impurity is $N_\mathrm{B}\approx60$. For the assumed imaging resolution, this corresponds to an optical depth $\sim1$ in the absorption image pixel containing the impurity, resulting in an easily observed transmission reduction of the imaging light, caused by the single impurity.

In conclusion, we have developed a novel method for spatially resolved imaging of single impurities in an ultracold gas. Our scheme provides access to direct study of impurity dynamics which can be straight-forwardly implemented in existing experiments.

The authors acknowledge support by EPRSC. B.O. also acknowledges funding by Fundaci\'on Ram\'on Areces. W.L. acknowledges funding by the EU. Discussions with C. Ates, T. Peyronel and H. Weimer are gratefully acknowledged.

During preparation of this manuscript we became aware of related work \cite{Gunter11}.

\end{document}